\documentclass[aps,prl,twocolumn,showpacs, nofootinbib,superscriptaddress]{revtex4-1}  
\usepackage{amsmath, amssymb,bbold}        
\usepackage{graphicx, multirow, color}
\usepackage{enumitem}
\setlist{leftmargin=5.5mm}
\usepackage{hyperref}

\newcommand{\MuMess}{\Lambda_{\text{mess}}}
\newcommand{\MuInt}{\Lambda_{\text{int}}}
\newcommand{\MuIR}{\mu_{\text{IR}}}

\newcommand{\gev}{~\text{GeV} }
\newcommand{\tev}{~\text{TeV} }
\newcommand{\strm}{\mathcal{S}}
\newcommand{\tInt}{t_{\text{int}}}
\newcommand{\aref}[1]{\hyperref[#1]{Appendix~\ref{#1}}}

\begin{document}

\begin{flushleft}
\hspace*{0.2cm}{TIFR/TH/19-8}
\end{flushleft}

\title{A radiatively generated source of flavor universal scalar soft masses}
\author{Sabyasachi Chakraborty}
\email{sabya@hep.fsu.edu}
\affiliation{Department of Physics, Florida State University, Tallahassee, FL 32306, USA,}
\author{Tuhin S. Roy}
\email{tuhin@theory.tifr.res.in}
\affiliation{Department of Theoretical Physics, Tata Institute of Fundamental Research, Mumbai 400005, India.}
\date{\today}
\begin{abstract}
We report that models of electroweak supersymmetry with gaugino mass unification and sequestered scalar masses  can still produce viable spectra,  as long as  we  include a set of non-standard supersymmetry breaking terms, which are trilinear in scalars like the A-terms, but are non-holomorphic in visible sector fields unlike the A- terms. These terms impart a subtle feature to one loop renormalisation group equations of soft supersymmetry breaking terms, indirectly sourcing flavor universal contributions to all scalar masses.  These new contributions can even dominate over radiative corrections from bino, and help raise right handed sleptons above bino, while leaving  a tell-tale signature in the spectrum.
\end{abstract}

\maketitle


The complete absence of any genuine hint of new physics from the LHC,  as well as null results in various direct and indirect searches for dark matter have put severe constraints on all models of electroweak (EW) supersymmetry. Probably nowhere  is this stress more visible than in models characterized by zero scalar masses at a  high scale. Often using ``locality" in set-ups with extra dimensions~\cite{Kaplan:1999ac, Chacko:1999mi}, or lattices of gauge groups connected by link fields~\cite{Csaki:2001em,Cheng:2001an},  or even strong and nearly conformal dynamics of the hidden sector~\cite{Roy:2007nz, Murayama:2007ge, Perez:2008ng}, these models sequester scalar masses at the input scale (say $\MuInt$), thereby ensuring that flavor universal gaugino mediation remains the sole source of scaler masses at infrared (IR). These elegant solutions to the  flavor problem in supersymmetry~\cite{Dimopoulos:1995ju} can also solve the $\mu$--$B_\mu$ problem~\cite{Roy:2007nz}, and provide a unique perspective to the fine-tuning problem~\cite{Perez:2008ng}, where dynamics brings in large cancellations in  Higgs mass matrix,  while at the same time, accommodating gaugino mass unification. Naturally, these models  provide examples of scenarios with maximum predictability~\cite{Perez:2008ng}, and not surprisingly, seem to be on the verge of being ruled out from cosmological considerations alone. The seed of this non-trivial claim lies in the fact that right handed (RH) sleptons, which only receive bino mediated contributions, are typically the lightest supersymmetric particles (LSPs) in these scenarios.  Consider, for example, the ratio $\tilde{m}^2_{E}/M_1^2$ determined at  $\MuIR$,  a scale in the IR: 
\begin{equation}
\frac{\tilde{m}^2_{E}}{M_1^2} \Big|_{\MuIR}  \simeq 
\frac{-6}{5b_1}
\left[1-\left\{1+\frac{b_1 g_1^2\left(\MuInt\right) } {8\pi^2}\ t_\text{int} \right\}^{2} \right] \; , 
\label{eq:right_slepton_mass}
\end{equation}      
where $\tilde{m}^2_{E}$ is the usual soft-squared mass for RH-sleptons, $M_1$ is the bino mass,  $b_1$ is the hypercharge beta function and  $\tInt$ is  $\log\left(\MuInt/\MuIR\right)$.  The R.H.S. of Eq.~\eqref{eq:right_slepton_mass}  is always less than one as long as $\MuInt \lesssim \ 4\times 10^{18}~\text{GeV} $. For any realistic spectra, therefore, bino is considerably heavier than RH-sleptons, and does not play any role in determining the nature of LSP. This immediately rules out the possibility of having a well-tempered neutralino to give thermal relic as cold dark matter~\cite{ArkaniHamed:2006mb}. The pure neutral wino or a Higgsino can give rise to the right relic for masses $\sim 2.5\tev$ or $\sim 1\tev$ respectively~\cite{Bramante:2015una}. However, given that  Fermi-LAT and HESS puts severe constraints on wino dark matter~\cite{Fan:2013faa, Cohen:2013ama}, higgsino seems to be the only safe candidate for a LSP. 

 Unfortunately, if one takes into account details for sequestering, which dictates that  $\MuInt$ is of  the order of the scale of supersymmetry breaking ( \textit{i.e.}, $ \lesssim 10^{10}$-$10^{11}\gev$)~\cite{Cohen:2006qc}, even the $\tev$ scale for LSP seems problematic. For  $\tilde{m}^2_E > \mu^2 \gtrsim \left( 1\tev\right)^2$, one needs $M_1 \gtrsim 2\tev$ at the EW scale. The further requirement of gaugino mass unification (\text{i.e.}, $M_3/g_3^2 = M_2/g_2^2 = M_1/g_1^2$) forces gluinos to be much heavier ($\sim 10\tev$). Obtaining electroweak symmetry breaking (EWSB) for such heavy gluinos while keeping $\mu$ at $1\tev$ is virtually impossible. 

The last remaining option is to have gravitino as the LSP. However, scenarios with a thermalized gravitino are riddled with issues. For example, the overabundance forces a gravitino mass bound  $m_{3/2} \lesssim 1~\text{keV}$~\cite{Pagels:1981ke}, whereas constraints from large scale formation  or the Tremaine-Gunn bound require $m_{3/2} > 0.4~\text{keV}$~\cite{Tremaine:1979we}. Note that such light $m_{3/2}$ arises when $\MuInt$  $\lesssim 10^{8}$-$10^{9}\gev$. In this case, Eq.~\eqref{eq:right_slepton_mass}  predicts $M_1 \gtrsim 3 \times \tilde{m}_E \gtrsim  1.3\tev$, where  we use $ \tilde{m}_E \gtrsim 450\gev$ --  the LHC bound on stable massive charged particles~\cite{Aaboud:2019trc,Khachatryan:2016sfv}, since RH-sleptons with these parameters are expected to be collider stable. Gravitino production through freeze out can  also yield the right relic~\cite{Bolz:2000fu}, but BBN constraints give an upper bound on the gravitino mass to be at or around $10\gev$~\cite{Kawasaki:2008qe}. These considerations become even more difficult if  hidden sector interactions are taken into account, which enhance  gravitino mass w.r.t. other superpartners because of renormalization~\cite{Roy:2007nz, Murayama:2007ge}.  In this work we do not consider the possibility of gravitino as the candidate of dark matter any more.  

The key realization in this letter is that most of these difficulties with  gaugino mediated spectra  (even after accommodating gaugino mass unification) can be avoided, if one manages to break the inequality $\tilde{m}^2_{E} \ll M_1^2$ at the EW scale. As explained before, this feature is generic as long as bino mediation remains the sole source of masses for RH-sleptons. On the other hand, if a new source  that can generate  $\tilde{m}^2_{E}$ radiatively  in a flavor universal way is identified, it  opens up the possibility of having $\tilde{m}^2_{E} \gtrsim M_1^2$.  In that scenario,  bino can be the LSP and a candidate for the dark matter with a correct thermal relic because of slepton coannihilation. The small splitting between bino and RH-sleptons required for coannihilation, implies that the decay of these sleptons to bino  yields only \emph{soft} leptons, thereby significantly loosening LHC bounds on slepton masses. New techniques involving soft muons~\cite{Sirunyan:2018iwl} and tracks~\cite{Chakraborty:2016qim,Chakraborty:2017kjq} can help unearth these spectra.

Finding a source of flavor universal masses for RH-sleptons is, however, highly non-trivial. A known source is the radiatively generated contributions from  the ``$\strm$-term"  defined as
\begin{equation}\label{eq:hypD}
\begin{split}
\mathcal S  \ \equiv \ & \text{Tr}\left(Y_\phi m^2_{\phi}\right)  \ =  \ \left(m^2_{H_u}-m^2_{H_d} \right)  \\ 
&  \qquad \  +  \text{Tr}\left(m^2_{Q}-m^2_L-2m^2_U+m^2_D+m^2_E\right) \; ,
\end{split}
\end{equation} 
where $\phi$ runs over all scalar particles with hypercharge  $Y_\phi$. However, in the present context this seemingly simple solution does not work since sequestering also ensures a zero $\strm$-term at $\MuInt$.  Also, since  $\left(\strm/g_1^2\right)$ is RGE invariant at one loop, $\strm$ remains zero even at the EW scale. Although, inhomogeneous pieces for $\strm$ are generated at three loops in the MSSM~\cite{Jack:1999zs} and are therefore too small to have observable consequences.
 
Before proceeding, we provide a brief understanding of these statements. First, consider renormalization below $\MuInt$ where only visible sector interactions matter. Note that a suitable field redefinition of the hypercharge vector superfield $V_Y$ can  absorb a non-zero $\strm$.
\begin{equation}
\begin{aligned}
V_Y \ \rightarrow \ V_Y + \kappa \theta^2 \bar{\theta}^2  \strm   
\qquad \text{implies} \\
\int d^4 \! \theta  \Phi^\dag e^{Y_\phi V_Y} \Phi \rightarrow 
\int d^4 \! \theta \Phi^\dag e^{Y_\phi V_Y} \Phi +  
\kappa Y_\phi \left| \phi\right|^2 \strm\;, \\ 
m^2_{\phi} \rightarrow  m^2_{\phi}  + \kappa  Y_\phi     \strm 
~~ \text{and}~~  \strm   \rightarrow   \strm 
+ \kappa       \strm \sum_\phi Y_\phi^2 n_\phi \; ,
\end{aligned}
\label{eq:DY-shift}
\end{equation}
where,  $n_\phi$ gives the dimension of the superfield $\Phi$ containing the scaler $\phi$.  Therefore, by choosing $\kappa$ judiciously one can make $\strm \rightarrow 0$. However this shift of $V_Y$ (or rather the shift of $D_Y$, the $D$-component  $V_Y$) does not disappear entirely from the Lagrangian.  The kinetic term for $V_Y$ also shifts, giving rise to  the Fayet-Iliopoulos (FI) term for hypercharge. 
\begin{equation}
\mathcal{L} \rightarrow \mathcal{L} + \int d^4 \theta \xi V_y 
\quad \text{with} \quad \xi = \frac{\strm}{g_1^2} \; .
\end{equation}
Since $\xi$ does not renormalize at one loop,  one immediately obtains the RGE for  $\strm$ in a familiar form
\begin{equation}
\frac{d}{dt} \left( \frac{\strm}{g_1^2} \right) = 0  \quad \Rightarrow 
\frac{d}{dt} \strm = \frac{b_1}{8\pi^2} g_1^2 \strm   \; .
\label{eq:s-rge}
\end{equation}
The $\strm$-RGE  in the presence of hidden sector couplings can be similarly calculated by  noting that $ m_\phi^2 = k_
\phi D^2 \bar{D}^2 \mathcal{R}/\MuMess^2$, where $D,  \bar{D}$ are superderivatives, $\mathcal{R}$ is the real operator, and $\MuMess$ is the scale of messengers. The shift in $V_Y$ in this case, is proportional to  $ D^2 \bar{D}^2 \mathcal{R}/\MuMess^2$. The resultant  FI-term generated is characterized by an operator, which runs because of anomalous dimension of $\mathcal{R}$ (say, $\gamma$) that also sequesters scalar masses.
\begin{equation} 
\xi  \ \propto \    \frac{1}{g_1^2} \frac{ D^2 \bar{D}^2 \mathcal{R}}{\MuMess^2}   \quad  \Rightarrow \quad  \frac{d}{dt} \xi  = \gamma  \xi \; .   
\label{eq:xi-hidden}
\end{equation}
Summarizing, in scalar sequestering $\strm$ is predicted to be zero at $\MuInt$ and it remains zero even at the IR. 
  
However, the  arguments presented above break down, if the set of supersymmetry breaking operators is expanded to include the  $C$-terms given by
\begin{equation}
C_t  Y_u \: h_d^\dagger \tilde{q} \tilde{u} \ + \ 
C_b  Y_d \: h_u^\dagger \tilde{q} \tilde{d} \ + \ 
C_\tau  Y_\tau \: h_u^\dagger \tilde{l} \tilde{e} \ + \ \text{h.c.}  \; .
\label{eq:Cterm}
\end{equation} 
The simplest way to understand this result involves analytically continuing to superspace, where the $C$-terms arise from supersymmetric operators containing, for example, $H_d^\dag \exp\left(V_Y /2\right) QU$. The factor $\exp\left(V_Y /2\right) $, the  presence of which is demanded by gauge invariance\footnote{We have suppressed weak gauge superfield in the exponent for simplicity.}, is not invariant under the shift in Eq.~\eqref{eq:DY-shift}. Therefore, in the presence of $C$-terms,  a theory with a non-zero $\strm$ is no longer equivalent to a theory with a nonzero $\xi$. 

Consequently, we expect the $\strm$-RGE to turn inhomogeneous even at one loop, with  $C$-terms acting as sources. This can be demonstrated diagrammatically~\cite{Jack:2000jr, Chakraborty:2018izc}.  Before proceeding, also note that we take these trilinear terms to be proportional to Yukawa couplings (same as the $A$-terms) in the spirit of minimal flavor violation~\cite{DAmbrosio:2002vsn}. Therefore, as long as we only consider the third generation Yukawa couplings (namely $y_t, y_b$ and $y_\tau$),  RGEs of the first two generations remain unaffected. The extra contributions  to the RGEs 
because of the $C$-terms  are given as 
\begin{equation}
\begin{split}
\delta \left( \frac{d}{dt}  \tilde{m}^2_{\phi} \right)  & =  \frac{2}{16 \pi^2}\ \sum_i \kappa_i^\phi \ \xi_i \; ,
 \qquad \text{where}  \\
&
 \xi_i  =  \left|y_i\right|^2 \left(\left|C_i + \mu \right|^2 - \left| \mu \right|^2 \right) \; ,
\label{eq:fields_RG}
\end{split}
\end{equation}
$\phi$ = $\left\{\tilde{Q}_3, \tilde{U}_3, \tilde{D}_3, \tilde{L}_3, \tilde{E}_3, H_u, H_d \right\}$,  the index $i$ runs over $\left\{t, b, \tau\right\}$, and    $\kappa$  is a matrix of numbers given as 
\begin{equation}
\begin{bmatrix} \kappa \end{bmatrix} \ = \ 
\begin{bmatrix}
\kappa_t \\
\kappa_b \\
\kappa_{\tau}
\end{bmatrix} \ = \ 
\begin{bmatrix}
1,2,0,0,0,0,3\\
1,0,2,0,0,3,0\\
0,0,0,1,2,1,0
\end{bmatrix} \; .
\end{equation}
To solve these we also need  RGEs for the $C$-terms:
\begin{equation}
\begin{split}
\frac{d}{dt} C_i \ &= \  \frac{1}{16\pi^2} \  \chi_{ij} C_j\;, 
\qquad  \text{where, } \\  
\begin{bmatrix}
\chi
\end{bmatrix} \ &= \ \omega  \mathbb{1}- 2 
\begin{bmatrix}
y_b^2 &  y_b^2 & 0\\
 y_t^2 &  y_{\tau}^2+y_t^2  & -y_{\tau}^2\\
0 & -3 y_b^2 & 3y_b^2
\end{bmatrix} \; ,
\end{split}
\label{eq:c-rge}
\end{equation}
and  $\omega=\left(3y_t^2+3y_b^2+y_{\tau}^2+3 g_2^2+3/5 \ g_1^2\right)$. The running of  gaugino masses, $A$-terms and the $\mu$-term remain unaltered and can be found in for example~\cite{Martin:1997ns}. 

Finally, using  the definition in Eq.~\eqref{eq:hypD}, we derive the $\strm$-RGE from Eq.~\eqref{eq:fields_RG} to be
\begin{equation}
16\pi^2 \frac{d}{dt}\strm  \ = \ \frac{66}{5} g_1^2 \strm - 12\ \xi_t +  12\ \xi_b + 4\ \xi_\tau  \; . 
\label{eq:srun}
\end{equation} 
As expected, we find inhomogeneous pieces in the RGE. This suggests that, contrary to common wisdom, $\strm$-term can be generated because of radiative corrections alone as long as $C$-terms are present at $\MuInt$. A non-zero $\strm$ contributes flavor universally to all scalars proportional to their hypercharges, and can potentially lift RH-sleptons above bino.  In order to get the main message through, we first solve the RGEs  neglecting $y_b$, $y_\tau, g_2$ and $M_1$ for simplicity.  The key equations to consider are:
\begin{align}
&\frac{d}{dt} C_t \ \approx \ \frac{3}{16\pi^2} \  y_t^2 C_t\;, 
\label{eq:Ct_sim}\\
&\frac{d}{dt}  \left(\frac{\mathcal{S}}{g_1^2} \right)\ \approx  -   \frac{3 }{4\pi^2} \frac{y_t^2}{g_1^2}\left(  C_t^2 + 2 \mu C_t \right) = -   \frac{3 }{4\pi^2}  \frac{\xi_t}{g_1^2}\;,  \\
&\frac{d}{dt}  \tilde m^2_{E}   \ \approx \ 
\frac{1}{16 \pi^2 } \  \frac{6}{5} \ g_1^2 \strm \; .
\label{eq:mer_sim} 
\end{align}
We find an approximate analytical solution for $\tilde{m}^2_{E_1}$ at $\MuIR$, given by (assuming $\tilde{m}^2_{E}= \strm = 0$ at $\MuInt$)
\begin{equation}
\frac{\tilde{m}^2_{E} \big|_{\MuIR}}{\xi_t/y_t^2\big|_{\MuInt}} \approx   \frac{2}{11}  \left(1 - G_1 \right) 
\left(\mathcal{Y} G_3^{8/3b_3} - 1 \right)\;,
\label{eq:slep_mass}
\end{equation}
where
\begin{equation}
G_a \ = \ \frac{\alpha_a\left(\MuIR\right)}{\alpha_a\left(\MuInt\right)} 
\quad \text{and} \quad  \mathcal{Y} \ = \  \frac{y_t\left(\MuIR\right)}{y_t\left(\MuInt\right)} \; .
\end{equation}
Using $G_1 < 1$ and $\mathcal{Y} G_3^{8/3b_3} < 1$ (valid in our regions of interest), we find that the input conditions  for $\xi_t$ should be negative in order to give positive definite $\tilde{m}^2_{E}$. For example,  $\xi_t \approx - \left(700\gev\right)^2$ is needed at  $\MuInt = 10^{11}\gev$ to generate a  $\tilde{m}^2_{E} \approx \left(100\gev\right)^2$ at $\MuIR=1\tev$. 

To be specific, we check the viability of our proposal in the context of scalar sequestering~\cite{Perez:2008ng}, an elegant version of conformal sequestering~\cite{Luty:2001jh, Luty:2001zv, Dine:2004dv, Ibe:2005pj, Ibe:2005qv, Schmaltz:2006qs}. This model is characterized by a zero $\strm$ at $\MuInt$, which remains the same at IR. To generate $\strm$ from running, we extend the set of operators to include non-zero $C$-terms at  $\MuInt$. The full set of initial condition at $\MuInt$ is specified as
\begin{equation}
\begin{gathered}
M_1, M_2, M_3  \qquad
\mu \qquad  A_t, A_b, A_{\tau} \qquad C_t, C_b, C_\tau\;,   \\
\tilde{m}^2_{H_u} \ =  \ \tilde{m}^2_{H_d} \ = \ -\left|\mu\right|^2\;,  \\
\tilde{m}^2_{Q}  = \tilde{m}^2_{U}  = \tilde{m}^2_{D}  = \tilde{m}^2_{L}  = \tilde{m}^2_{E}  =  0\;,  \quad  B_{\mu}=0\; .
\label{eq:Newseq}
\end{gathered}
\end{equation}
The number of free parameters can be reduced further, if one assumes gaugino mass unification.   

The physics of RH slepton masses in the first two generations (namely, $\tilde{m}_{E_{1,2}}^2$)  is the same as before. The boundary conditions in Eq.~\eqref{eq:Newseq}, however, give rise to an interesting effect when we consider RH-stau mass $\tilde{m}^2_{E_3}$.  Even if we start with the same boundary condition, $\tilde{m}^2_{E_3}$  runs differently than $\tilde{m}^2_{E_{1,2}}$ because of $y_\tau$.  In the limit $C_\tau \rightarrow 0$ but $y_\tau \neq 0$, it is straightforward to see that $\tilde{m}^2_{E_3} > \tilde{m}^2_{E_{1,2}} $ in the IR. This feature arises because of the negative definite Higgs soft masses at the boundary, and is rarely seen in typical MSSM scenarios.  In fact,  the splitting $ \Delta_{E}^2 \equiv  \tilde{m}^2_{E_3} - \tilde{m}^2_{E_{1,2}}$ remains positive throughout running, and gives a tell-tale signature of the boundary conditions in  Eq.~\eqref{eq:Newseq}. The size of the splitting, however, depends on $\tan \beta$ through $y_\tau$, and should increase with increasing $\tan \beta$. 
\begin{figure}[h!]
	\centering
	\includegraphics[height=4.1cm,width=.5\textwidth]{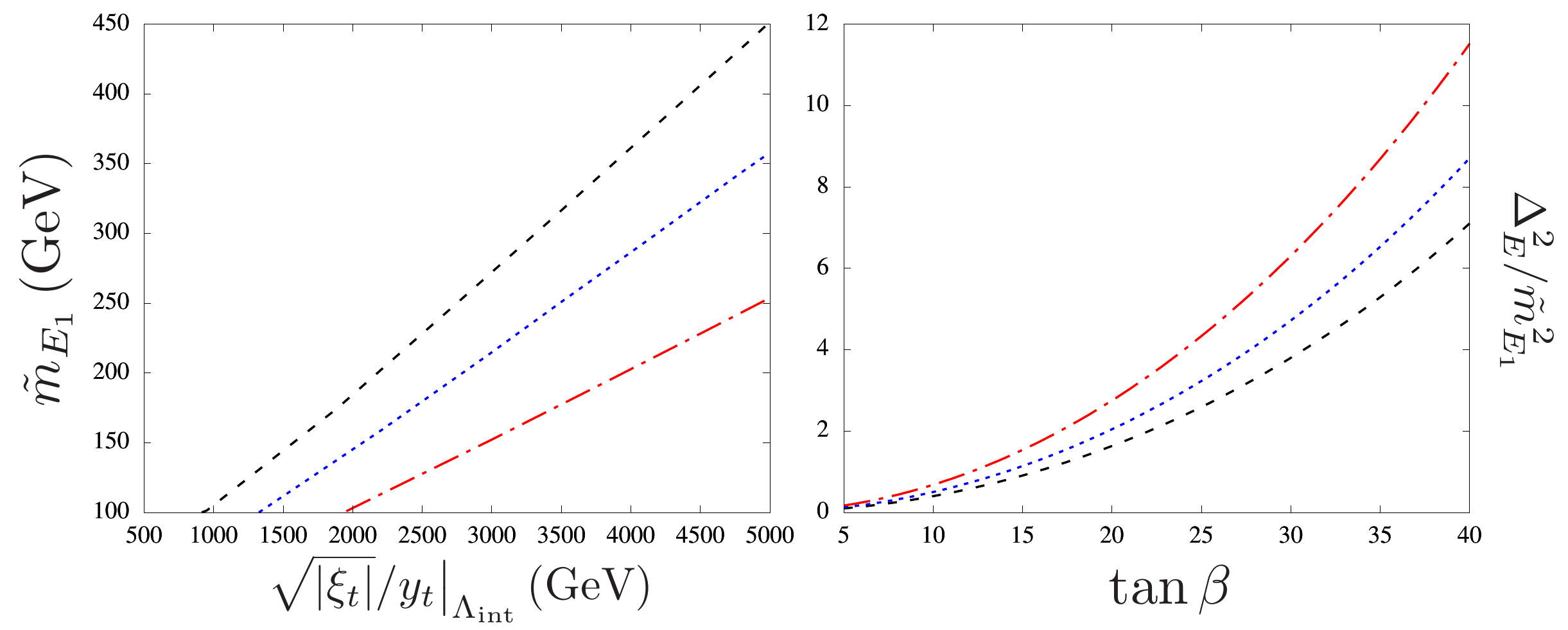}
	\caption{Left panel: first generation slepton mass as function of the input parameter $\sqrt{\left|\xi_t\right|}/y_t$ for different values of $\MuInt$ (black \& dashed: $10^{11}$ GeV, blue \& dotted: $10^9$ GeV, red \& dash-dotted: $10^7$ GeV) with $M_1=0$. Right panel: The mass splitting between the first and third generation RH sleptons as a function of $\tan\beta$.}
	\label{fig:RHslepton_seq}
\end{figure}
We confirm this behavior in Figure~\ref{fig:RHslepton_seq}. In the left plot we numerically solve the full set of RGEs (neglecting $M_1$) and show the variation of first two generation slepton masses with the initial condition $\sqrt{\left|\xi_t\right|}/y_t$, for different values of $\MuInt$.  In the right panel we show the variation of the  fractional splitting, namely $\Delta_{E}^2/\tilde{m}^2_{E_1}$ with   $\tan \beta$, keeping all other conditions unaltered. Not surprisingly, we find  splitting to increase with $\tan \beta$.     

\begin{figure}[h!]
	\centering
	\includegraphics[width=0.47\textwidth]{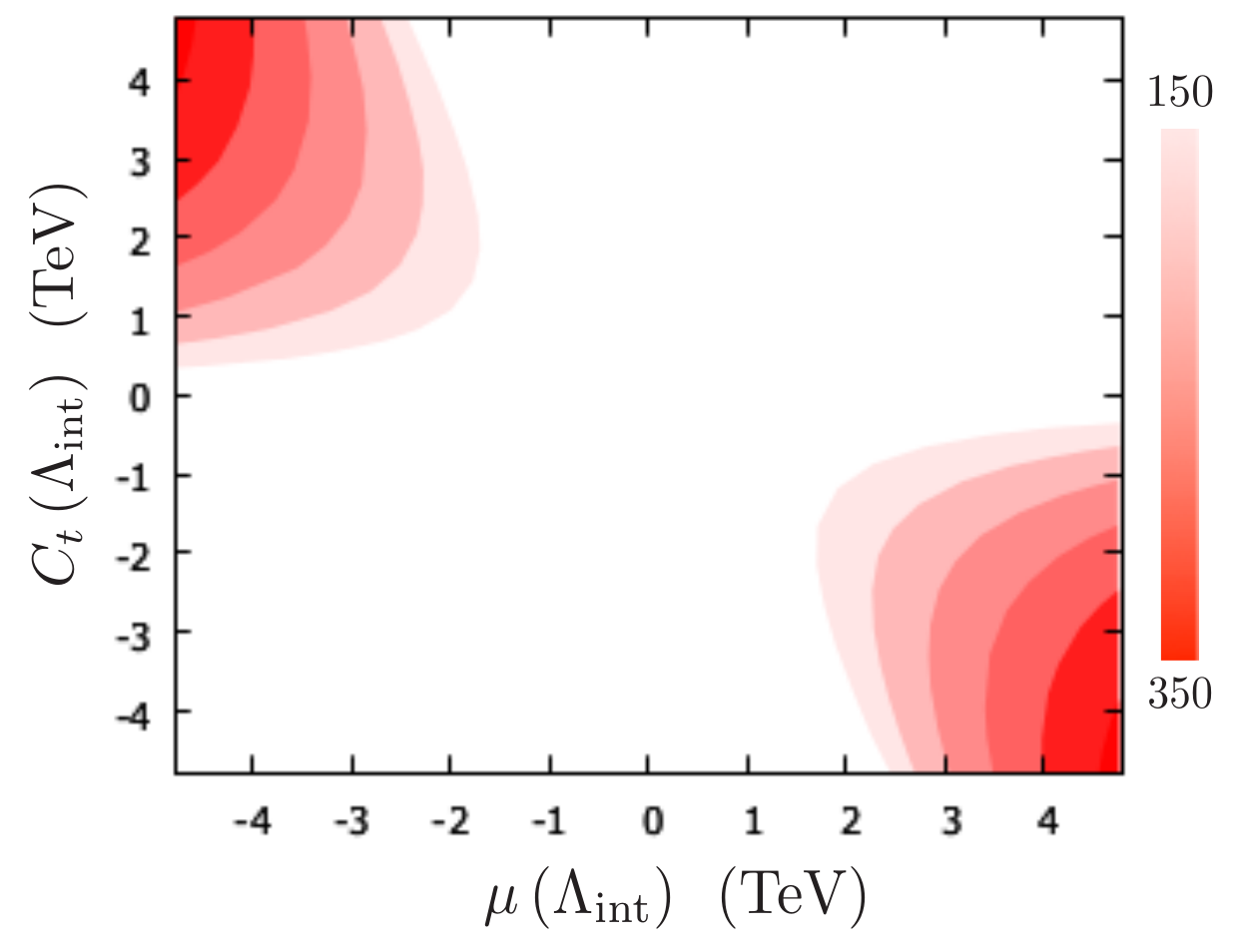}
	\caption{We show the contours of the first two generation slepton masses at $\MuIR$ (in GeV) in the $\mu$--$C_t$ plane fixed at the input scale. The allowed region requires $\mu$ and $C_t$ to have a relative sign in between them which consequently generates non-tachyonic slepton masses at $\MuIR$.}
	\label{fig:RHslepton_seq}
\end{figure}
As shown in Eq.~(\ref{eq:slep_mass}), $\xi_t$ is required to be negative in order to generate non-tachyonic mass for the right chiral sleptons. This immediately implies from Eq.~(\ref{eq:fields_RG}) that $C_t$ and $\mu$ needs to have a relative sign between them, as shown in Fig.~\ref{fig:RHslepton_seq}. Out of these two choices, negative $\mu$ is more appealing because such a choice is less constrained by dark matter direct detection searches~\cite{Grothaus:2012js}. Even though using a large $C_t$ at $\MuInt$ one can make RH-sleptons substantially heavier than bino (the LSP), cases where these states are nearly degenerate have many advantages. As mentioned before, these spectra allow for a LSP with the correct thermal relic because of bino-slepton coannihilation~\cite{Griest:1990kh,Jungman:1995df,Edsjo:1997bg,Ellis:1998kh,Desai:2014uha}.  As long as the mass difference of bino and the RH sleptons is of the order of the freezeout temperature $T_F$, all these states are  thermally accessible and are nearly as abundant.  Considering $T_F \sim M_1/25$ one finds that coannihilation is active as long as   $\left(\tilde{m}_{E} -M_1\right)/M_1 \sim 0.05$~\cite{Griest:1990kh}. The relic, however, also depends on $M_1$, and overabundance assuming standard cosmology, imposes a constraint   $M_1 \sim \tilde{m}_{E} < 400\gev$. Additionally, in the limit of small $\tan \beta$, RH-stau is almost degenerate with other RH-sleptons. Consequently, the scale  $M_1 \sim \tilde{m}_{E}$ can be as low as $\sim 100\gev$, since the compression among bino and RH-sleptons allows us to weaken collider bounds~\cite{Aaboud:2017leg, Sirunyan:2017lae,Aaboud:2018jiw}.

Further difficulties arise if gauginos are assumed to be on the unification trajectory. Using previous bound on $M_1$,  we find that  $M_3 =  \left( \alpha_3/\alpha_1       \right) M_1 \lesssim 5 \times 400\gev$ at the EW scale. This runs into issues from direct LHC bounds on gluinos~\cite{Aaboud:2017vwy,Sirunyan:2017kqq}. Moreover, since squark masses get sizable contributions only from gluino mediation, we find $M_3 > 3\tev$ from squark bounds. This issue is also raised in Ref.~\cite{Martin:2017vlf}, which found that a thermal relic with gauginos in the unification trajectory can not be reconciled even if one uses pseudo-fixed points as initial conditions instead of vanishing scalar masses. 

Note, however,  that this problem can be easily resolved  when the reheating temperature $T_R$  is lower than $T_F$.  The crucial consideration is that the reheating mechanism, when entropy gets continuously injected due to the decay of inflaton/moduli fields, is \emph{not} taken to be an instantaneous process.  Expressed in terms of the Hubble parameter, the universe during reheating expands according to $H\propto T^4$, as opposed to $T^2$ in radiation dominated universe, or $T^{3/2}$ in matter dominated universe. The relic abundance (determined at $T_F$)   naturally gets diluted due to the relatively faster expansion of the universe. Simply following the prescription as chalked out in Ref.~\cite{Giudice:2000ex}, one can start with a seemingly over-abundant thermal relic (\textit{i.e.} a large $M_1$), and can still produce the right abundance today given the right $T_R$.  
\begin{figure}[t]
	\centering
	\includegraphics[width=0.45\textwidth]{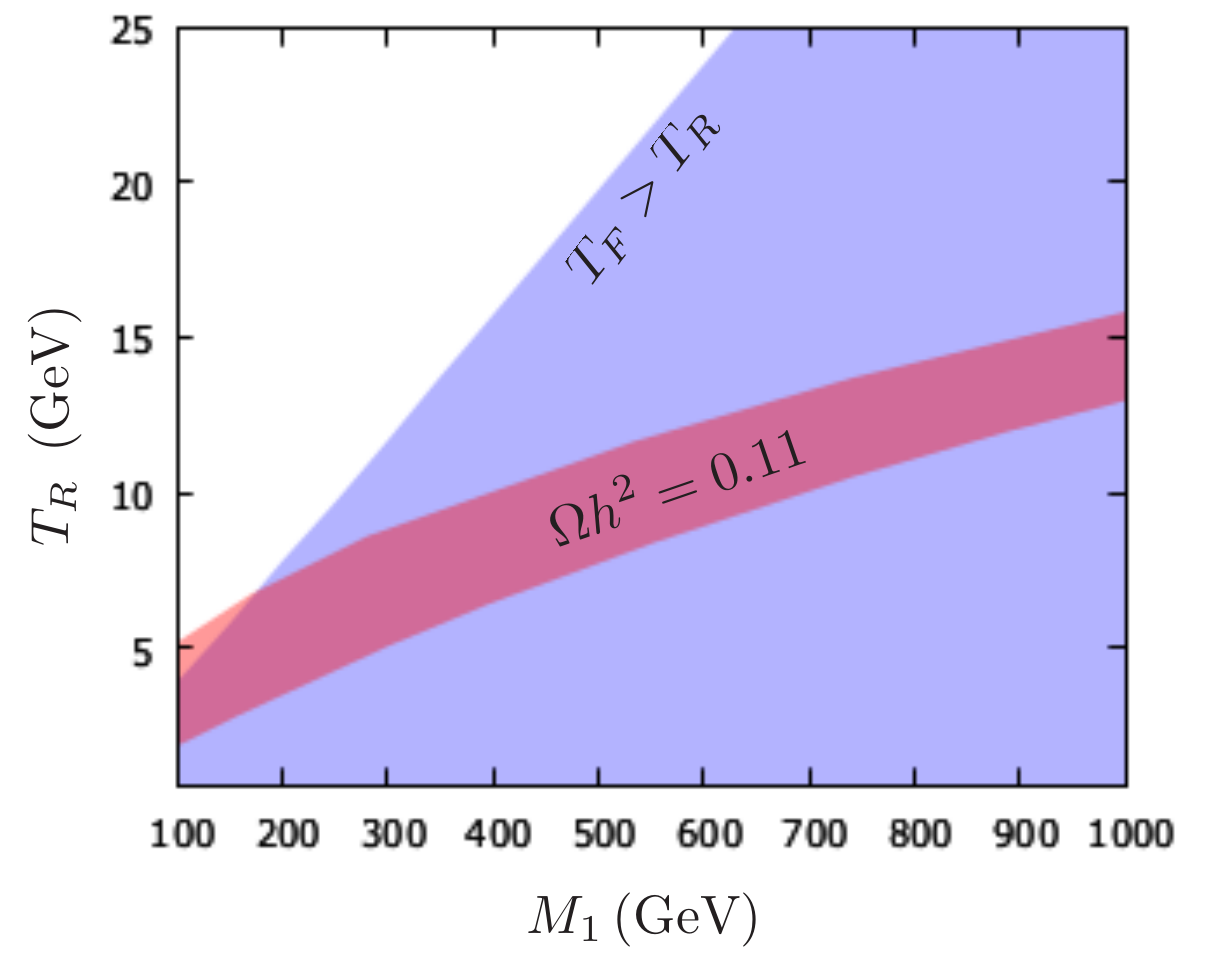}
	\caption{The red shaded region satisfies the relic density $\Omega h^2=0.11$~\cite{Aghanim:2018eyx}. The upper and lower boundaries correspond to boundaries in $20\gev \leq \left( \tilde m_E - M_1 \right) \leq 500\gev$. The blue shaded region refers to the condition where $T_F>T_{R}$.}
	\label{fig:8}
\end{figure}
In Fig.~\ref{fig:8}, we show the allowed region satisfying the relic density in red in the $M_1$--$T_R$ plane. This patch corresponds to different values of the right slepton mass with the imposed constraint that $20\gev \leq \left( \tilde m_E - M_1 \right) \leq 500\gev$. The region where $T_F>T_R$ is shown in blue. The region where these two patches do not overlap is obviously excluded. We observe that even large masses for the LSPs  are compatible, if the reheating temperature is small (\textit{i.e.}, in the GeV range). 
\begingroup
\squeezetable
\begin{table}[h!]
	\centering
	\begin{tabular}{|c | c | c |} 
		\hline
		Parameters ($\Lambda_{\text{int}}$) & BP1 (GeV) & BP2 (GeV) \\ [0.5ex] 
		\hline
		$M_{3,2,1}$ & 2000, 1000, 281    & 1850, 1300, 846\\
		$C_{u,\tau}$ & 1030, -350   & 6000, 2000 \\
		$A_t,\mu$ & 115, -2500  & 2200,-5500 \\ 
		$\tan\beta$ & 10 & 7\\
		\hline
		\hline
		$M_{3,2}$ & 3680, 904  & 3404, 1175 \\
		$\tilde m_{\tilde q_{1}, \tilde u_1, \tilde d_1}$ & 2960, 2910, 2915  & 2788, 2666, 2712\\
		$\tilde m_{\tilde t_1, \tilde b_1, \tilde L_1}$ & 2873,  2925, 511 & 3566,  2733, 589 \\
		$\tilde m_{\tilde t_2, \tilde b_2, \tilde L_2}$ &2995, 2960, 528 & 4122, 3567, 676\\
		$\mu$ & -2485  & -5514  \\
		$\underline{M_1, \tilde m_{E_{1,3}}}$ & 192, $\left(201,212\right)$ & 579.5, $\left(584,593\right)$ \\
		$m_{A,h}$ & 621,126  & 3882,126 \\
		$\Omega h^2$ & 0.117 & 0.48 \\
		$\sigma_{\text{SI}}(\text{cm}^2)$ & $5\times 10^{-48}$ & $4.6\times 10^{-48}$ \\
		\hline
	\end{tabular}
	\caption{Benchmark points for scalar sequestering with $C$-terms. The top (bottom) panel shows parameters at $\MuInt$ ($\MuIR$). UV parameters not mentioned here are kept at zero.}
	\label{table:1}
\end{table}
\endgroup

As concrete evidence that the mechanism proposed here can result in viable spectra, we provide two benchmark points In Table~\ref{table:1}. One of these points additionally assumes gaugino mass unification (\textbf{BP2}). In order to calculate the spectra, we set $\MuInt = 10^{11}\gev$ and $\MuIR = 1\tev$. For calculating relic density and direct detection rates we use \texttt{MicrOMEGAs5.0}~\cite{Belanger:2018ccd}. In \textbf{BP1},  gaugino masses can be set independently, which  allows us to have the LSP (and RH-sleptons) as light as $200\gev$. On the other hand,  we need to invoke a low  $T_F \sim 10\gev$ in  \textbf{BP2} to get the right thermal relic.  In both the benchmarks, the direct detection cross-sections are below the existing bound of Ref.~\cite{Aprile:2018dbl}.  Note EW symmetry breaking (EWSB) is highly nontrivial given the boundary conditions in Eq.~\eqref{eq:Newseq}. Unlike standard MSSM scenarios, $\mu$ and $B_\mu$ are input conditions and can not be derived at the EW scale from the EWSB equalities.  Both the benchmark points given in this work satisfy EWSB conditions. 

Even though References~\cite{Ellwanger:1983mg, Bagger:1993ji, Jack:1999ud, Martin:1999hc, Jack:1999fa, Hetherington:2001bk, Jack:2004dv, Nelson:2015cea, Sabanci:2008qp, Chakrabortty:2011zz, Martin:2015eca, Ross:2016pml, Chattopadhyay:2016ivr, Ross:2017kjc, Chattopadhyay:2017qvh, Beuria:2017gtf, Chattopadhyay:2018tqv}  that  have explored phenomenological consequences of the $C$-term are aplenty, in this work  we, for the first time, find its usage in building models of EW supersymmetry. These terms, if present, impart a subtle feature to the RGEs of soft supersymmetry breaking terms, which allows us to resurrect an elegant class of models.  Generating these terms, though, is highly nontrivial. The model in Ref.~\cite{Nelson:2015cea},  which is shown to be equivalent to a theory with the $C$-terms~\cite{Chakraborty:2018izc}, provides one such example. Another way to generate such terms could be conceived in models of gauge mediated supersymmetry breaking where some of the messenger fields couple to the MSSM Higgs field~\cite{Haber:2007dj}. Deriving the RGEs of these $C$-terms in the presence of arbitrary supersymmetry breaking dynamics, or to build models where $C$-terms can flow to nontrivial  fixed points in the context of scalar sequestering, however, remain  open questions, which we leave for future endeavor.

\section*{acknowledgements}
We thank Takemichi Okui and Adam Martin for their early readings and comments on the manuscript. SC is supported by the US Department of Energy under grant DE-SC0010102. TSR was supported in part by the Early Career Research Award by Science and Engineering Research Board, Dept. of Science and Technology, Govt. of India (grant no. ECR/2015/000196). We also acknowledge the workshop ``Beyond the Standard Model: where do we go from here?'' hosted at the Galileo Galilei Institute for Theoretical Physics (GGI), as well as the workshop ``International Meeting on High Energy Physics" hosted at Institute of Physics (IOP) where parts of this work was completed. 

\bibliography{References.bib}
\end{document}